\documentclass[multphys,vecphys]{svmult}

\usepackage{makeidx}         
\usepackage{graphicx}        
\usepackage{multicol}        
\usepackage[bottom]{footmisc}

\makeindex             

\begin{document}

\title*{The Formation Histories of Metal-Rich and Metal-Poor
Globular Clusters}
\author{Stephen E. Zepf}
\institute
{Dept. of Physics and Astronomy, Michigan State University,
\texttt{zepf@pa.msu.edu}}
%
%
\maketitle

\vspace{-12pt}
\paragraph{Abstract}

This review presents the results of ongoing studies of the 
formation histories of metal-poor and metal-rich globular 
clusters and their host galaxies. I first discuss the strong
observational evidence that the globular cluster systems of
most elliptical galaxies have bimodal metallicity distributions. 
I then focus on new results for metal-poor and metal-rich globular 
cluster systems. Metal-poor globular clusters are often associated
with early structure formation, and I review new constraints
on their formation epoch based on the ``bias'' of the number 
of metal-poor clusters with host galaxy mass. For metal-rich
globular clusters, I discuss new results from ongoing optical
to near-infrared photometric studies which both confirm an
intermediate-age population in NGC~4365 and generally reveal
a variety of formation histories for now quiescent ellipticals.


\vspace{-8pt}
\section{Introduction}
As noted by several speakers at this meeting, many of us begin 
our talks and proposals giving reasons why we study globular 
clusters (GCs) to learn about the formation of their host galaxies. 
However, even if standard, listing these reasons is important because
the characteristics that make globular clusters useful for studying
the formation of their host galaxies are often revisited when
comparing the results of the many different programs aimed at
constraining galaxy formation and evolution. In this spirit,
the following are some of the key reasons globular clusters are 
valuable tools for understanding the formation history of their 
host galaxies.

1) Because the age and metallicity can be determined for each globular
cluster individually, the distribution of ages and metallicities
within a galaxy population can be constrained.

2) Globular clusters are the best example we have of a simple 
stellar population, so determining the age and metallicity of
individual GCs is much simpler than studies of integrated light
in which the stars of many different metallicities and ages are 
seen as one luminosity weighted average.

3) Globular clusters are observed to form in all major star formation
events in galaxies, making them tracers of the major formation episodes 
of their host galaxies. 

4) Some globular clusters are among the oldest known objects, so
they may provide constraints on early structure formation.

5) As dense concentrations of $\sim 10^6$ stars, globular clusters
can be observed in galaxies across the local universe. This enables 
the formation history of a representative sample of galaxies to be 
studied, including those of normal elliptical galaxies which are not 
present at very nearby distances.

In this review, I will focus on two new results regarding the
formation history of globular cluster systems (GCSs) and their host
galaxies. One of these new results is a constraint on the formation
epoch of metal-poor globular clusters from the determination of 
their cosmological ``bias'' with host galaxy mass 
(see also Rhode, Zepf, \& Santos 2005, and Katherine Rhode's talk 
at this meeting). The second set of results are constraints on
the formation history of elliptical galaxies from the age
distributions of their globular cluster systems. I specifically
review new work on the optical to near-infrared colors of
globular cluster systems, which includes deep NICMOS
data confirming previously identified intermediate age
populations in the giant elliptical galaxy NGC~4365, and 
new PANIC data showing an effectively completely old
population in another giant elliptical, NGC~4472.
I then discuss the overall status of work on ages, and
look to future samples and comparisons with elliptical galaxy
formation histories estimated in other ways.

\vspace{-8pt}
\section{A Note about Metallicity and Color Distributions}

The primary results discussed below involve using metal-poor
globular clusters to probe early structure formation and metal-rich
ones to probe the formation epochs of their host galaxies.
While clearly some globular clusters have higher or lower
metallicities than others, it is still natural to consider the 
basis for dividing globular cluster systems into metal-poor 
and metal-rich populations.

The obvious observational starting point is that most globular
cluster systems of elliptical galaxies are observed to have
bimodal color distributions. This was first noted many years ago
now (Zepf \& Ashman 1993) and much subsequent work has made it
clear that the globular cluster systems of most, although not all, 
elliptical galaxies have bimodal color distributions
(e.g.\ Kundu \& Whitmore 2001, Larsen et al.\ 2001).
It was also realized early on that for most elliptical galaxies
GCSs, the optical color primarily traces metallicity. 
This is because metallicity is the 
primary driver for optical colors at ages greater than a few Gyr,
and elliptical galaxies generally do not have huge amounts of
recent star formation. It turns out the minority of ellipticals
which have unimodal GCS color distribution may be an interesting 
exception to this rule as several of these have been found to 
have younger GC populations (see Section 4), but for typical 
bimodal systems, the GC color primarily traces metallicity.

Given that GC color primarily traces metallicity in typical
bimodal systems, the next question is what is the detailed relationship 
between these two, and specifically whether the bimodal color
distribution observed for most elliptical galaxy GCSs
reflects a truly bimodal metallicity distribution.
One common way to determine the relationship between color and
metallicity is to fit the observed relation for the Milky Way GCS.
Data for Galactic globular clusters is fundamental for this question,
because only for these GCs are there true abundances from
high resolution spectra for individual stars, as well as 
detailed color-magnitude diagrams for age and metallicity 
determination. This approach to the relationship between color 
and metallicity is then limited by the Galactic globular cluster 
sample. Its primary limitation for this purpose is the absence of
high metallicity clusters. Additional concerns are that some
of the few with higher metallicities also have large and
uncertain reddenings, that the Galactic GCs do not extend
to young ages, and simply the modest total number of Galactic
GCs.

Another common approach is to use stellar populations models.
However, these are not a panacea, because they are tested on
Galactic data, and therefore have many of the same concerns
at old ages and high metallicities as empirical fits of the 
Galactic GCS color-metallicity relation. Another possibility
is using absorption-line indices for extragalactic systems
to investigate the behavior of colors at higher metallicities.
However, these indices are only related to actual abundances
through stellar populations models or through Galactic GCs,
and are therefore also strongly dependent on uncertainties 
in the models and any gaps in the Galactic sample on 
which these are tested.

There is a large body of work on determining color-metallicity
relations using the above approaches and applying these to 
extragalactic globular cluster systems. These cover many
color indices, including B$-$I and C$-$T1 which have large wavelength 
baselines and for which there are extensive datasets, and V$-$I for
which there is significant HST archival data.
The conclusion of these many studies is that while the exact
shape of the metallicity distribution varies somewhat 
depending on the color-metallicity relation, the general
bimodal nature of the metallicity distribution remains. It is 
very difficult to create a strong dip right in 
middle of the color distribution without a bimodal-like metallicity
distribution. A variety of kinematical studies also find 
differences between the metal-poor and metal-rich globular 
cluster systems (e.g.\ Zepf et al.\ 2000, C\^ot\'e et al.\ 2003), 
providing independent evidence for the presence of these two 
subpopulations, similar to the
way in which kinematics provides further evidence for two 
populations of GCs in the Milky Way.
One contrary claim has recently been put forward and
was presented at this meeting by Yoon et al.\ (see these proceedings).
They use a stellar populations model with a specific treatment
of the horizontal branch, and show their model gives a very
sharp feature in the color-metallicity relation at the precise
place to mimic a metallicity bimodality.
They also claim that this sharp feature is consistent
with the preliminary g$-$z Galactic GC data in Peng et al.\ (2006),
although no comparison is made to other well-established
Galactic GC colors with a similar wavelength baseline.

A powerful argument that the observed color bimodality is
due to a truly bimodal metallicity distribution is that the 
Galactic GCS is known without question to be bimodal in metallicity.
While one might say we happen to live in a galaxy with a 
complicated history, it would seem quite strange for us to 
have ended up in a complicated galaxy, while other galaxies 
have the virtue of uncomplicated pasts. This ``Copernican''
argument for metallicity bimodality is strongly supported
by a number of other observational results.

One observation that provides a clear indication
of a bimodal metallicity distribution is bimodality in
the I$-$H color distribution for the M87 GC system 
(Kundu et al.\ 2006). The near-infrared I$-$H color is dependent
almost completely on metallicity, and it is very hard to
image how a horizontal branch effect as proposed by Yoon et al.\
can account for bimodality at these near-infrared wavelengths. 
This evidence for bimodality in metallicity is supported
by a wide variety of other data. Perhaps the simplest is the 
analysis of many different colors vs metallicities for Galactic
GCs, including the B$-$I and C$-$T1 shown in talks at this meeting, 
both of which have broad wavelength ranges, does not show the very 
sharp and precisely located jump in color vs.\ metallicity required 
to produce the observed bimodal color distribution from a unimodal 
metallicity distribution. As noted in Smits et al.\ (2006),
even linear fits of these Galactic GC color-metallicity
relations do not have much more scatter than the observational
uncertainties. Cohen et al.\ (2003) also noted in their fits
of color to Galactic GC metallicities and absorption-line indices
for extragalactic systems that a second-order fit is only slightly
better than a linear fit, and their Figs. 13 and 15 do not
provide evidence for sharp jumps in the color-metallicity relation.
In this context is it worth noting that most absorption line
studies of extragalactic GCSs are not fair samples of the cluster
metallicity distribution, as many intentionally select their 
GC targets to have colors in the middle of the distribution
in an attempt to decrease contamination of the sample by
non-clusters. Therefore, while the comparison of colors and
absorption-line indices for individual objects is valuable (e.g.\ Cohen
et al.\ 2003), the distribution of absorption-line indices 
in extragalactic samples is not generally useful because
of the preferential selection of GCs with intermediate colors.

Thus, there is both direct evidence for bimodal metallicity 
distributions as described above and the simple argument that 
it seems unlikely we are privileged to live in one of the few 
galaxies that has GC system with a unimodal metallicity distribution.
These strongly indicate that the bimodal color distributions 
observed for GC systems arise from bimodal metallicity distributions.
Therefore, for the remainder of
this review, we will consider the metal-poor and metal-rich
GC populations separately.

\vspace{-8pt}
\section{The Formation Epoch of Metal-Poor Globular Clusters
from Their Bias with Galaxy Mass}

It is natural to associate metal-poor globular clusters
with early epochs of structure formation when the overall enrichment
in galaxies was not high. Moreover, for many Galactic globular clusters,
age determinations also indicate an early epoch of formation, although
the uncertainties in GC ages even for Galactic clusters are large when
converted to redshift in the early universe. Therefore, an independent
method to assess the formation epoch of metal-poor globular clusters
would be very interesting.

One idea for constraining the general formation epoch of the metal-poor
globular cluster population is to take advantage of the feature of
hierarchical structure formation models that more massive objects
have a greater fraction of their mass collapsed at higher redshift,
and that this ``bias'' increases fairly steeply at very high $z$.
This has recently been pursued by Rhode, Zepf, \& Santos (2005),
based in part on the discussion in Santos (2003), and is also 
discussed in this meeting in Katherine Rhode's talk. Specifically,
we used new wide-field multi-color imaging for both ellipticals 
and spirals in a range of environments to determine the total 
number of metal-poor globular clusters around these galaxies. 
We then determined the galaxy-mass normalized number of metal-poor
globular clusters for each galaxy (their ``T$_{blue}$'' value, and
plotted this versus galaxy mass. If the the metal-poor globular
clusters formed at moderate redshift when most of the mass has
collapsed for typical galaxies, there would be little or no bias 
and all galaxies would have similar T$_{blue}$ values regardless 
of galaxy mass. With an extremely high redshift of formation for 
the metal-poor globular clusters, the mass-normalized number of 
metal-poor GCs would show a steep rise to higher galaxy masses, 
as only the most massive galaxies had even a small fraction of 
their mass collapse at very early epochs.

Our results are that the mass-normalized number of metal-poor GCs 
(T$_{blue}$) increases with increasing galaxy mass (Rhode et al.\ 2005).
However, the increase is not particularly steep, suggesting extremely
high redshifts for the typical formation epoch of metal-poor globular
clusters are unlikely (see Katherine Rhode's talk for relevant plots). 
Moreover, most systematics tend to steepen the relation, hinting that 
the current result may be an upper limit to the cosmological ``bias'' 
of the metal-poor GC population and its formation redshift. Further
work in this area will include more detailed modeling of the theoretical
expectations (see Moore et al.\ 2006 for such effort), use of the 
radial profile of the GCSs as a constraint, and of 
course needed increases in the size of the galaxy sample with data 
that provides sufficient areal coverage and depth to obtain a reliable 
determination of the total number of metal-poor GCs.

\vspace{-8pt}
\section{The Formation Epoch(s) of Elliptical Galaxies
from the Ages of Their Metal-Rich Globular Clusters}

Determining the formation history of massive early-type galaxies
is one of the primary challenges of current extragalactic astronomy.
These elliptical and S0 galaxies make up a significant fraction
of the stellar mass in the local universe, but there is not yet a
consensus on when they formed. Because the age of individual
globular clusters can be determined, and globular cluster formation 
is observed to be a ubiquitous feature of starbursts, determining
the age distribution of their globular clusters is a valuable
way to constrain the formation history of elliptical galaxies.

There are several different observational approaches for determining
the age and metallicity of extragalactic GCs. Although I have
been involved in some way with nearly all of them, in this review
I will focus on the optical to near-infrared color approach, as
it is both very promising and not the subject of other talks
at this meeting. The optical to near-infrared color technique
is very promising because it relies on straightforward stellar
physics to determine the age and metallicity, and because it
is a photometric technique and therefore observationally efficient.
The technique solves the age-metallicity degeneracy
for unresolved simple stellar populations because the infrared
color (e.g.\ I$-$H) is primarily sensitive to metallicity, while
the optical color (e.g.\ g$-$I) has a greater sensitivity to
age. Thus in plots of optical to near-infrared colors age and 
metallicity are separated (see Fig.\ 1). This separation is about 0.3
mag between intermediate ($\sim 3$ Gyr) and old ($\sim 15$) Gyr.
A comparable example from another area is that 0.3 mag
is similar to the difference in SN Ia brightnesses in an
accelerating universe compared to an open one.

\begin{figure}
\centering
\includegraphics[angle=-90,width=11cm]{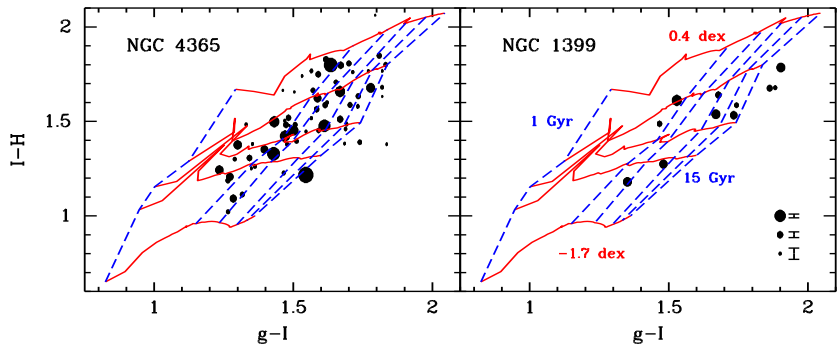}
\caption{The g$-$I vs.\ I$-$H plots of globular clusters in NGC~4365 and 
NGC~1399 from NICMOS and ACS data from Kundu et al.\ (2005). 
These show a substantial population of
GCs with optical to near-ir colors indicative of intermediate age
in NGC~4365, in agreement with earlier work (P02), while NGC~1399 
has few such GCs. Only GCs with uncertainties less than 
0.1 mag in each axis are shown. The lines trace age 
(1, 3, 5, 8, 11, and 15 Gyr from the left) and 
metallicity ([Fe/H] = $-1.7, -0.7, -0.4, 0.0,$ and 0.4 dex from the bottom) 
contours from BC03 models. The size of the points are inversely proportional 
to the I$-$H uncertainty as shown at the bottom right. The g$-$I 
uncertainties are comparable to the I$-$H values in NGC~4365 and are 
much smaller than the I$-$H uncertainties in NGC~1399.}
\label{fig:1}
\end{figure}

One of the first results to come from the application of the optical
to near-infrared technique was the discovery of a substantial
population of intermediate age globular clusters in the elliptical
galaxy NGC~4365 (Puzia et al.\ 2002, hereafter P02). This result
was originally thought to be surprising, in part because
spectroscopy of the integrated light of NGC~4365 was originally 
thought to indicate an old age and thus presented a puzzle.
However, recent calculations using alpha-enhanced isochrones 
necessary for NGC~4365 and its large observed [Mg/Fe] give younger 
ages (Brodie et al.\ 2005, hereafter B05). Extant spectroscopic
studies of the globular clusters themselves are inconclusive 
as one finds intermediate ages and the other does not, even for
objects in common between the two studies (B05).

Therefore, to pin down the ages and metallicities of the GCs
around NGC~4365, we obtained new, very deep NICMOS near-infrared
photometry in three fields, and combine this with new ACS data.
As shown in Figure 1, these independent and much higher signal-to-noise
data confirm the previous result that NGC~4365 has a substantial 
population of GCs with optical to near-infrared colors that
can only be accounted for by intermediate ages of $2-7$ Gyr
(Kundu et al.\ 2005). We also show in this paper that the
result is independent of which stellar population models are 
used, although the exact age of the intermediate age population 
does vary with model. In addition, Kundu et al.\ (2005) used 
archival data for NGC~1399 to show that its GCS is predominantly 
old with a small younger component, in agreement with previous 
spectroscopic work.

A comparison of optical to near-infrared
photometry and the absorption-line spectroscopy for different
galaxy GCSs reveals generally good agreement (see Kundu et al.\ 2005). 
Another example shown in Figure 2 is new PANIC data we have obtained 
for NGC~4472, which shows an almost exclusively old GC system 
(Hempel et al.\ 2006). This is in agreement with extant spectroscopic 
studies (Cohen et al.\ 2003, Beasley et al.\ 2000).
The key advantages of the optical to near-infrared approach 
are that as a photometric technique it is more efficient to
obtain age and metallicity estimates of a given uncertainty
in the same amount of observing time, and that it relies on
straightforward stellar astrophysics. The advantages of
absorption-line indices are that they enable estimates of 
abundance ratios, which are interesting probes of timescales
of formation (e.g.\ Pierce et al.\ 2006).

\begin{figure}
\centering
\includegraphics[width=8.5cm]{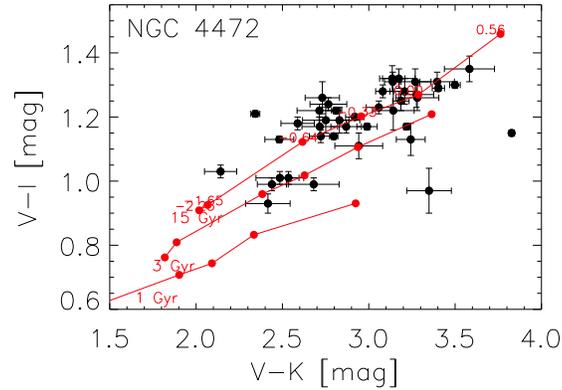}
\caption{The V$-K$, V$-$I plots for globular clusters around NGC~4472,
indicating that the GCS of NGC~4472 appears to be nearly completely old.
In this plot, the model ages decrease left to right, going from 
15 Gyr on the upper left to 3 Gyr on the lower right, and metallicity 
increases from [Fe/H] = $-2.2$ in the lower left part to 0.5 on the
upper right part of the plot. All of these are BC03 models, and the data
are from our ongoing work (Hempel et al.\ 2006).}
\label{fig:1}
\end{figure}
%
%

The work described here has been made possible by my many excellent
collaborators. I acknowledge support for this research from NSF award 
AST-0406891, NASA LTSA grant NAG5-11319, STScI grant 
HST-GO-09878.01-A, and a NSF travel grant.


%
%
%
%
%
%
%
%
%

\vspace{-8pt}
{}
\printindex
\end{document}